\documentclass[times,twocolumn,final]{elsarticle}
\usepackage{framed,multirow}

\usepackage{amssymb}
\usepackage{latexsym}

\usepackage{url}
\usepackage{xcolor}

\usepackage{adjustbox}
\usepackage{multirow}
\usepackage{booktabs}
\usepackage{amsmath}
\usepackage{hyperref}
\hypersetup{pdfauthor=whatever}
\usepackage{amsfonts}
\usepackage{amssymb}
\usepackage{pifont}% http://ctan.org/pkg/pifont
\usepackage{colortbl}
\usepackage{caption}
\usepackage{enumitem}

\definecolor{cgreen}{rgb}{0,0.7,0.6}
\journal{Arxiv}

\begin{document}

\title{Towards Universal Unsupervised Anomaly Detection in Medical Imaging}%

\author[1,2]{Cosmin I. Bercea\corref{cor1}}
\cortext[cor1]{Corresponding author: cosmin.bercea@tum.de}
\author[3]{Benedikt Wiestler}
\author[1,3,4]{Daniel Rueckert}
%% Third author's email
\author[1,2,5]{Julia A. Schnabel}

\address[1]{Technical University of Munich, Munich, Germany}
\address[2]{Helmholtz AI and Helmholtz Center Munich, Munich, Germany}
\address[3]{Klinikum Rechts der Isar, Munich, Germany}
\address[4]{Imperial College London, London, UK}
\address[5]{King's College London, London, UK}

\begin{abstract}
%%%
The increasing complexity of medical imaging data underscores the need for advanced anomaly detection methods to automatically identify diverse pathologies. Current methods face challenges in capturing the broad spectrum of anomalies, often limiting their use to specific lesion types in brain scans. To address this challenge, we introduce a novel unsupervised approach, termed \textit{Reversed Auto-Encoders (RA)}, designed to create realistic pseudo-healthy reconstructions that enable the detection of a wider range of pathologies. We evaluate the proposed method across various imaging modalities, including magnetic resonance imaging (MRI) of the brain, pediatric wrist X-ray, and chest X-ray, and demonstrate superior performance in detecting anomalies compared to existing state-of-the-art methods. Our unsupervised anomaly detection approach may enhance diagnostic accuracy in medical imaging by identifying a broader range of unknown pathologies. Our code is publicly available at: \url{https://github.com/ci-ber/RA}.
%%%%
\end{abstract}

\begin{keyword}
%% Keywords
Generative AI \sep Unsupervised Anomaly Detection \sep Medical Imaging
\end{keyword}

\maketitle

\begin{figure*}[tb]
    \includegraphics[width=\textwidth]{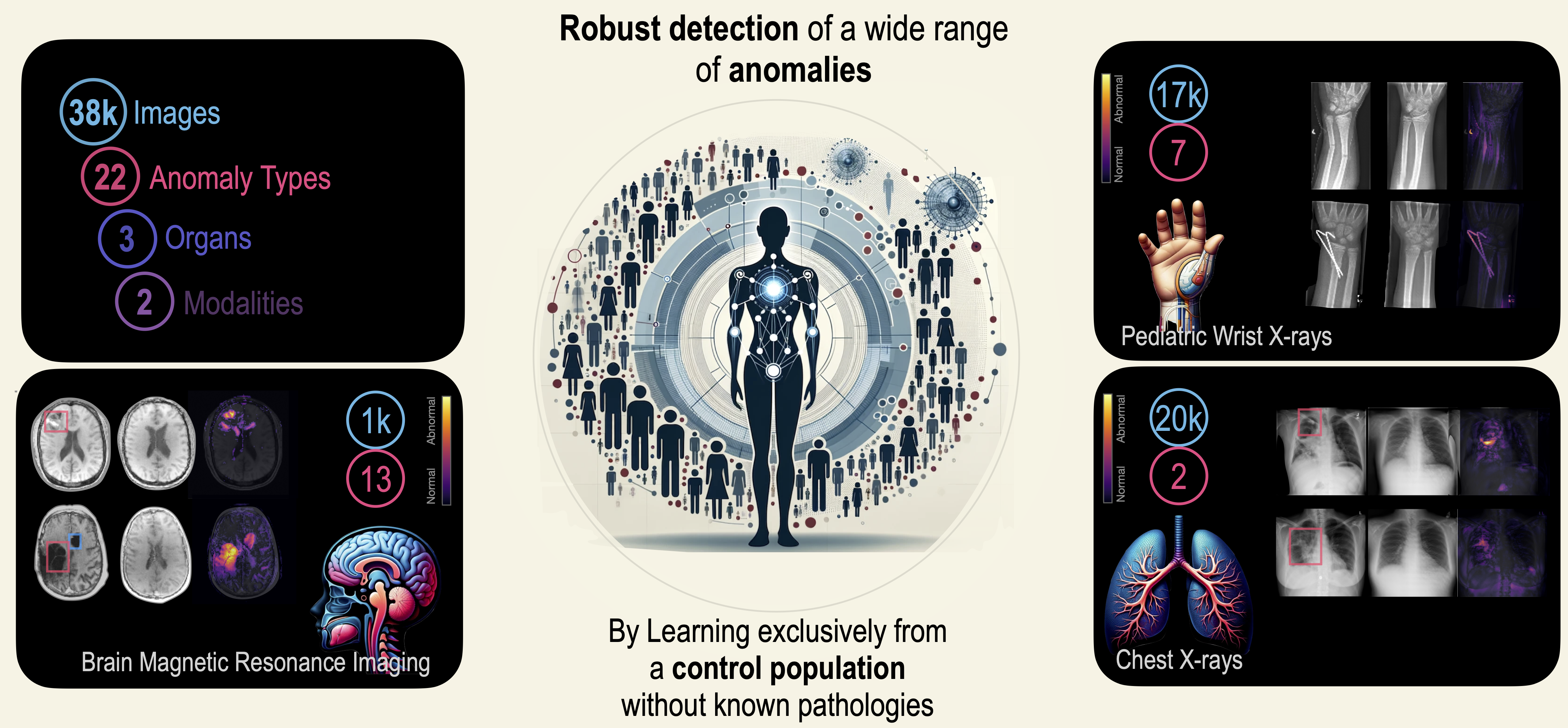}
    \caption{\textbf{Towards Universal Unsupervised Anomaly Detection.} The figure illustrates the detection of various anomalies in a dataset comprised of $\approx$ 38,000 images, spanning 22 anomaly classes, 3 anatomies, and 2 imaging modalities. The representation underscores the capacity of the model to learn from normal populations and its effectiveness in identifying unknown anomalies highlighting its potential utility in clinical screening and diagnostic applications.\label{fig::teaser}}
\end{figure*}

\section{Introduction}

Imaging is integral to diagnosis, treatment decisions, and disease monitoring in medicine. The rapid advancements in imaging technology have led to an exponential increase in both the volume and complexity of imaging data, necessitating more sophisticated methods for analysis~(\url{https://data.oecd.org/healthcare/magnetic-resonance-imaging-mri-exams.htm}). Anomaly detection has emerged as a crucial technique for identifying abnormal patterns or structures, highlighting the underlying pathologies, and thereby assisting in the critical step of pathology detection in the diagnostic cascade.

Historically, anomaly detection in medical imaging has relied heavily on supervised methods, designed to identify specific, well-defined pathologies like tumors~\citep{menze2014multimodal}, stroke~\citep{atlas2022}, or white matter hyperintensities~\citep{kuijf2019standardized}. While effective in these specific scenarios, these methods inherently suffer from biases towards the expected anomaly distributions and are constrained in their applicability beyond the specific pathologies they are designed to detect. This limitation has significant implications, as it narrows the scope of detectable pathologies and overlooks a broad spectrum of potential anomalies in medical imaging.

Unsupervised anomaly detection (UAD) offers a promising alternative, aiming to detect anomalies without reliance on predefined labels. However, a significant challenge in UAD has been its tendency to focus on evaluations using singular or a limited number of conditions, or employ self or weakly supervised methods~\citep{wolleb2022diffusion,kascenas2022denoising} to estimate the 'unknown' in anomaly detection. This can potentially compromise the fundamental principle of unsupervised learning, which is to detect anomalies in a broad, unbiased manner.

Given that disease (anomaly) detection is the central first step in the diagnostic process and represents a significant source of error in radiology~\citep{kim2014fool}, the importance of developing unbiased anomaly detection methods is clear. Our approach to unbiased anomaly detection proposes a novel generative AI method, trained exclusively on normal anatomical samples. This approach is designed to restore pseudo-healthy versions of pathological inputs, thereby facilitating a nuanced and universal detection of anomalies. We have rigorously evaluatedted our method across a diverse range of modalities, as shown in~\autoref{fig::teaser}, including brain MRI, pediatric wrist X-ray, and chest X-ray images. The results consistently demonstrate the effectiveness of the proposed method in detecting a wide array of anomalies, across different anatomies and imaging techniques. Our approach signifies a substantial step forward in the field of unsupervised anomaly detection in medical imaging, offering a more accurate, unbiased, and comprehensive tool for medical professionals. In summary, our contributions include:
\begin{itemize}
    \item \textbf{Introduction of \textit{Reversed Auto-Encoders (RA)}:} We propose a novel generative AI methodology, termed \textit{'Reversed Auto-Encoders' (RA)}, designed to reconstruct pseudo-healthy versions of pathological inputs.
    \item \textbf{Extensive Evaluation of State-of-the-Art Methods:} Our study comprehensively evaluates various state-of-the-art (SOTA) anomaly detection methods across a spectrum of pathologies, anatomies, and modalities.
    \item \textbf{High Accuracy and Robustness for Anomaly Detection:} The RA method consistently outperforms existing SOTA methods in detecting anomalies across all tested pathologies, anatomies, and imaging modalities. This underlines the robustness and effectiveness of RA in a wide range of clinical scenarios, contributing significantly to the advancement of anomaly detection techniques.
\end{itemize}
\section{Anomaly Detection in Medical Imaging}
Anomaly detection in medical imaging is fundamentally concerned with unveiling the unknown—a statistical process aimed at identifying deviations from established normative patterns. It operates on the principle of outlier detection, where data points that significantly differ from the majority of a dataset are flagged as anomalies. Anomaly detection algorithms must be designed to be agnostic to specific pathologies, capable of generalizing across diverse data sets, and proficient at discerning unseen and varied anomalies. This requires algorithms to be trained on broad datasets, encompassing a wide range of normal variations, to effectively identify outliers without reliance on labeled data for specific conditions.

\textbf{Self-supervised methods}~\citep{self-cvpr-2021,self-tmi-2021,self-eccv-2022,self-kainz-2022,self-jiang-2023} leverage data augmentation or pretext tasks to generate surrogate supervisory signals. These methods exploit data-intrinsic features and limited annotations to discern anomalies, which promises a detection mechanism that can adapt to unseen anomalies. However, they may inadvertently instill bias in the expected anomaly distribution, notably when noise or artificial alterations serve as proxies for genuine pathological features. For example, Denoising Autoencoders (DAEs)~\citep{kascenas2022denoising} propose to learn to eliminate synthetically added coarse Gaussian noise. While this approach may demonstrate promising results for specific anomalies such as brain tumors, it is limited in its applicability to more general anomaly detection scenarios, as it relies on the distribution of the synthetic noise~\citep{bercea2022ra}. Weakly-supervised methods~\citep{weak-mickael-2021,weakWang2021,weak-ke-2022,weakHibi2023-pk,weak-dauchelle2023}, on the other hand, utilize partial or noisy labels to guide the anomaly detection process. For instance,~\cite{wolleb2022diffusion} proposed a diffusion-based anomaly detection method that utilizes guidance from a supervised classifier trained specifically for brain tumor segmentation. While this approach achieves promising results in detecting brain tumors, it inherently relies on the performance of the supervised classifier and its ability to provide effective guidance during the diffusion process. As a consequence, its applicability to more general anomaly detection tasks may be limited.

\textbf{Unsupervised anomaly detection} aims to learn the normative distribution from a normal population and subsequently apply this knowledge to anomaly detection. This approach necessitates a robust understanding of what constitutes 'normal' in medical images as a baseline for recognizing deviations.

Knowledge distillation has emerged as a pivotal technique in this context, facilitating the transfer of complex patterns and insights from complex models trained on extensive datasets to simpler models trained on a normal data subset. This enables the detection of anomalies by capitalizing on the discrepancies between the predictions of the teacher (larger model) and the student (simpler model)~\citep{salehi2021multiresolution,bergmann2020uninformed}. However, adapting this technique to the intricate and high-dimensional nature of medical imaging datasets poses significant challenges~\citep{bercea2022ra}. 

Traditional Autoencoders (AEs) have been fundamental in establishing reconstruction-based methods for anomaly detection~\citep{zimmerer2018context}. Employing an encoding-decoding architecture, AEs aim to capture and reconstruct input data, hypothesizing that anomalies will manifest as significant reconstruction errors. However, AEs often struggle to learn detailed normal anatomy features while not generalizing well to pathologies~\citep{bercea2023aes}. Variational AEs (VAEs)~\citep{kingma2013auto} have significantly contributed to advancing anomaly detection by addressing some of the limitations inherent in traditional AEs. This is achieved by regularizing the latent space and conceptualizing it as a probabilistic distribution. Such regularization allows for a more constrained learning process, enabling VAEs to adhere more closely to the normative distribution. This adherence is crucial in medical imaging, where the precise characterization of normal anatomy is imperative for effective anomaly detection~\citep{zimmerer2019unsupervised}. However, while beneficial, the regularization frequently results in the generation of blurrier reconstructions. This blurriness can be a drawback when fine details are critical for identifying subtle anomalies~\citep{bercea2023aes}. 

Likelihood models focus on characterizing the likelihood of normal data, and evaluating how well new samples conform to the learned normal distribution. A pivotal advancement in this domain has been the introduction of normalizing flows~\citep{kobyzev2020normalizing}. These provide a refined mechanism for transforming simpler probability distributions into more intricate ones, enhancing the precision in estimating data sample likelihoods. However, when applied to the high-dimensional and intricate nature of medical imaging data, normalizing flows encounter challenges, particularly in maintaining the accuracy of the reconstructions~\citep{zhao2023aeflow}. Latent Transformer Models (LTMs) have emerged as a notable innovation within likelihood models~\citep{PINAYA20221}. LTMs incorporate transformer networks within the latent space of a model to effectively identify and modify potentially anomalous instances. 

Masked AEs (MAEs) also capitalize on the strengths of advanced neural network architectures, but they approach the problem of anomaly detection from a different angle. MAEs utilize a strategy of selectively masking portions of the input data and tasking the model with predicting these occluded sections. By predicting the masked parts of an image, MAEs essentially learn a comprehensive representation of normal anatomy~\citep{he2022masked,schwartz2022maeday,lang20233d}.

Generative Adversarial Networks (GANs) have introduced adversarial training methodologies that have enabled the generation of highly realistic images, marking a new epoch in image synthesis and anomaly detection capabilities~\citep{goodfellow2014generative,schlegl2019fanogan}. However, they may suffer from mode collapse or may generate images not representative of the input data. To address these challenges, advancements like Soft-Introspective VAEs (SI-VAEs) have emerged~\citep{daniel2021soft}. They fuse VAEs and GANs and aim to overcome the specific limitations of GANs in anomaly detection. 

Deviating from the reliance on constrained latent spaces, Denoising Diffusion Probabilistic Models (DDPMs) employ an iterative methodology involving the addition and subsequent removal of noise directly in the image space~\citep{ho2020denoising}.  However, a critical aspect of DDPMs lies in the careful selection of noise levels, a decision that greatly influences their performance~\citep{graham2022denoising,bercea2023mask}.

Collectively, these developments mark significant advancements in anomaly detection in medical imaging. However, their evaluation has often been limited to narrow datasets, which may not fully represent the vast gamut of anomalies encountered in medical practice. This limitation raises questions about the universality and overall performance of the SOTA methods in broader, more diverse clinical scenarios. To address this gap, we extensively evaluate various cutting-edge methods (including \textit{RA}) using a comprehensive benchmark dataset. This benchmark encompasses a wide range of diseases, anatomies, and imaging modalities, thus providing a more rigorous and holistic assessment of their capabilities in universal anomaly detection.
\section{Background}
\begin{figure*}[tb]
    \includegraphics[width=\textwidth]{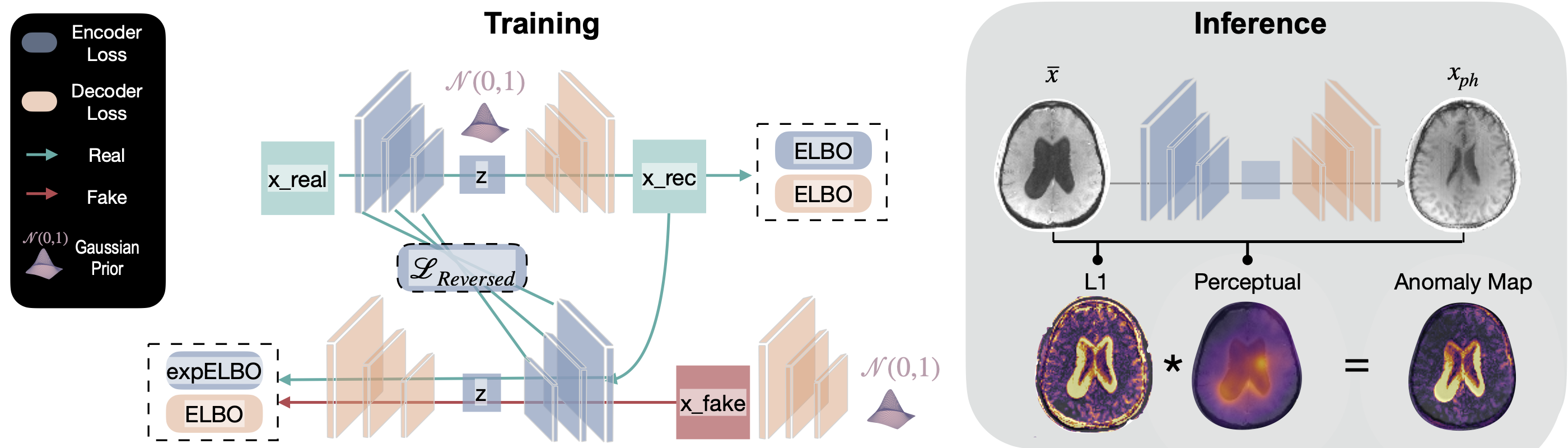}
    \caption{\textit{Reversed Autoencoder (RA)} framework during training and inference phases. During the training phase (left), the encoder and decoder networks are optimized using a multi-scale reversed embedding loss \(\mathcal{L}_{\text{Reversed}}\), in conjunction with the Evidence Lower Bound (ELBO) and adversarial optimization. In this process, the decoder generates a synthetic image \(x_{\text{fake}}\) from random noise, with the goal of fooling the encoder into treating it as a real image. In the inference phase (right), the RA model processes a new input \(x\), encoding and reconstructing it into a pseudo-healthy image \(x_{\text{ph}}\). Anomaly detection is carried out by computing the L1 norm and perceptual differences between \(x\) and \(x_{\text{ph}}\), resulting in an anomaly map that highlights pathological regions. \label{fig::arch}}
\end{figure*}

In the context of UAD, we refer to 'normal' as the absence of pathologies. Given a set of normal samples \(x \in \mathcal{X} \subset \mathbb{R}^N\), the objective of AEs is to find functions \(f:\mathbb{R}^N\rightarrow\mathbb{R}^D\) and \(g:\mathbb{R}^D\rightarrow\mathbb{R}^N\) such that \(x\approx g(f(x))\). Typically, \(f\) and \(g\) are referred to as the encoder and decoder, respectively, with \(f\) mapping the input to a lower-dimensional representation. The fundamental assumption in UAD is that these learned representations contain features describing the normative distribution, even for outlier samples \(\overline{x} \notin \mathcal{X}\)~\citep{bercea2023aes}. Consequently, \(x_{ph}=(g(f(\overline{x}))) \in \mathcal{X}\) represents the pseudo-healthy reconstruction of \(\overline{x}\). An anomaly score is usually derived from the pixel-wise difference between an input and its reconstruction: \(s(x) = |x-g(f(x))|\).

Within the variational inference framework~\citep{kingma2013auto}, the goal is to optimize the parameters \(\theta\) of a latent variable model \(p_{\theta}(x)\) by maximizing the log-likelihood, \(\log p_{\theta}(x)\), of the observed samples \(x\). However, the likelihood term is often intractable. To address this, the true posterior \(p_{\theta}(z|x)\) is approximated by a proposal distribution \(q_{\phi}(z|x)\) using the Evidence Lower Bound (ELBO):
\begin{equation}
    \label{eq::elbo}
    \log p_\theta(x) \geq \mathbb{E}_{q(z|x)}[\log p_\theta(x|z)] - \text{KL}[q_\phi(z|x) || p(z)] = \text{ELBO}(x)
\end{equation}
Here, KL denotes the Kullback-Leibler divergence; \(q_\phi(z|x)\) and \(p_\theta(x|z)\) are the encoder \(E_\phi\) and decoder \(D_\theta\), neural networks with parameters \(\phi\) and \(\theta\). VAEs often use a normal distribution \(p(z) = \mathcal{N}(\mu, \sigma)\) as the prior and employ the reparameterization trick to maximize the ELBO.

To combine the latent properties of VAEs with the image synthesis abilities of GANs, SI-VAEs~\citep{daniel2021soft} introduce an adversarial loss to the VAE training. The key innovation is to utilize the VAE's encoder and decoder in an adversarial manner, without the need for external discriminators. The encoder aims to differentiate between real and generated samples by minimizing the KL divergence of the latent distribution of real samples and the prior while maximizing it for generated samples. Conversely, the decoder is trained to 'fool' the encoder by reconstructing real data samples using the standard ELBO and minimizing the KL divergence of generated samples compressed by the encoder. The optimization objectives for the encoder and decoder are as follows:
\begin{flalign}
    \label{eq::sivae}
    & \mathcal{L}_{E_{\phi}}(x,z) = \text{ELBO}(x) - \frac{1}{\alpha}(\exp(\alpha \text{ELBO}(D_\theta(z)))),\\
    & \mathcal{L}_{D_{\theta}}(x,z) = \text{ELBO}(x) + \gamma \text{ELBO}(D_\theta(z)), \nonumber
\end{flalign}
where \(\alpha \geq 0\) and \(\gamma \geq 0\) are hyperparameters.

\section{RA: Reversed Autoencoders}
To advance the field of anomaly detection in medical imaging, we introduce the reversed AEs \textit{(RA)}. The primary innovation of \textit{RA} lies in its sophisticated training mechanism, designed to learn and accurately reconstruct normal anatomical patterns, a critical aspect for effectively distinguishing pathologies (see~\autoref{fig::arch}). This is achieved through a unique combination of three distinct training strategies. Firstly, the ELBO is employed to regularize a smooth latent space, enabling the model to effectively capture the underlying distribution of normal anatomical features. Secondly, an introspective adversarial interplay between the encoder and decoder components of the \textit{RA} is implemented. This interplay ensures the generation of high-fidelity representations of the normative distribution, as the encoder and decoder challenge each other to refine their outputs. Finally, to enhance the coherence between the input and its reconstruction - particularly critical in the restoration phase where substantial divergence can occur - we introduce a 'reversed loss'. This loss function is designed to minimize discrepancies between the original image and its reconstructed version, thereby ensuring that the \textit{RA} maintains a high degree of accuracy in reconstructing normal anatomy while simultaneously highlighting anomalies.

\subsection{Reversed Embedding Similarity}
Central to our approach is the implementation of a reversed multi-scale embedding similarity loss within the encoder. This methodology ensures close alignment of input representations with the embeddings of their generated reconstructions, performed at multiple scales:
\begin{align}
    \mathcal{L}_{Reversed}(x) = \sum_{l=0}^L \bigg[ & (1 - \mathcal{L}_{Sim}(E_\phi^l(x), E_\phi^l(x_{rec}))) \nonumber \\
    & + \frac{1}{2} \text{MSE}(E_\phi^l(x), E_\phi^l(x_{rec})) \bigg],
    \label{eq::reversed}
\end{align}
\begin{table*}[t]
\caption{Performance Metrics of Anomaly Detection Methods Across Varied Medical Conditions. Cells colored in \colorbox{red!20}{red} indicate detection rates below 50\%; \colorbox{yellow!20}{yellow} highlights rates below 60\%; and \colorbox{cgreen!20}{green} denotes rates above 60\%. The best results are emphasized in \textbf{bold}, while the second-best results are \underline{underlined}. Notably, \textit{Reversed Auto-Encoders (RA)} consistently exhibit commendable performance across all diseases, achieving the highest F1 score. This table underscores the importance of diverse benchmarks, revealing disparities such as DAE's proficiency in edema detection contrasted with its limitations in identifying enlarged ventricles and craniotomy. Visual results from \textit{RA} are presented in~\autoref{fig::adult_brain}.\label{tab::fast}}
    % \caption{Performance metrics of anomaly detection methods across varied medical conditions. Cells marked in \colorbox{red!20}{$red$} denote detection rates below 50\%; \colorbox{yellow!20} {yellow} indicates rates below 60\%; and \colorbox{cgreen!20}{green} signifies rates above 60\%. Best results are marked in {\boldmath $bold$}, and second-best are \underline{underlined}. Notably, only \textit{RA} exhibited commendable performance across all diseases with the highest F1 score. The table highlights the value of diverse benchmarks, spotlighting disparities like DAE's strength in edema detection versus its challenges with enlarged ventricles and craniatomy. Visual results of \textit{RA} are shown in~\autoref{fig::adult_brain}.\label{tab::fast}}
    \centering
    \setlength{\tabcolsep}{4pt}
        \begin{adjustbox}{width=\linewidth,center} 
            \centering
            \begin{tabular}{l  l | cc || cc | cc | cc | cc | cc | cc | cc}
                \toprule	    
                & \multirow{2}{*}{Method} & \multicolumn{2}{c||}{Total} & \multicolumn{2}{c|}{Edema} & \multicolumn{2}{c|}{Mass} & \multicolumn{2}{c|}{Lesions} &  \multicolumn{2}{c|}{Resection} & \multicolumn{2}{c|}{Enlarged Ventricles} & \multicolumn{2}{c|}{Craniatomy} & \multicolumn{2}{c}{Absent Septum} \\
                & & \#det & {F1 $\uparrow$}& \#det & {F1 $\uparrow$}& \#det & {F1 $\uparrow$}& \#det & {F1 $\uparrow$}& \#det & {F1 $\uparrow$}& \#det & {F1 $\uparrow$}& \#det & {F1 $\uparrow$}& \#det & {F1 $\uparrow$} \\\midrule

               %%%%% DAE %%%%%
                 \textit{Self-S}& DAE~\cite{kascenas2022denoising}  & 102/171 & \underline{$31.52$}& {\cellcolor{cgreen!20} \boldmath $17/18$} & {\cellcolor{cgreen!20} \boldmath$64.72$}& {\cellcolor{cgreen!20} \underline{19/26}} & {\cellcolor{cgreen!20} \underline{$25.68$}} & {\cellcolor{cgreen!20} \boldmath $19/22$} & {\cellcolor{cgreen!20} \boldmath $47.11$}& {\cellcolor{yellow!20}6/10} & {\cellcolor{yellow!20}\underline{$44.39$}}  & {\cellcolor{red!20}9/19} & {\cellcolor{red!20}$35.09$} &{\cellcolor{red!20} 6/15} & {\cellcolor{red!20}$16.29$} & {\cellcolor{red!20}0/1}& {\cellcolor{red!20}$0.00$}\\
                \midrule
            
          	     \multirow{7}{*}{\rotatebox[]{90}{\textit{Unsupervised}}} & MKD~\cite{salehi2021multiresolution} & $87/171$ & $26.61$ & {\cellcolor{cgreen!20}\underline{$14/18$}} & {\cellcolor{cgreen!20}$42.5$} &  {\cellcolor{yellow!20}$15/26$} & {\cellcolor{yellow!20}$15.48$} & {\cellcolor{yellow!20}11/22} & {\cellcolor{yellow!20}21.96} & {\cellcolor{cgreen!20}7/10} & {\cellcolor{cgreen!20}38.54} & {\cellcolor{cgreen!20}\underline{17/19}} & {\cellcolor{cgreen!20}\boldmath$80.12$} &{\cellcolor{red!20} 0/15} & {\cellcolor{red!20}0.00} & {\cellcolor{cgreen!20}\boldmath$1/1$} & {\cellcolor{cgreen!20}\boldmath$50.00$} \\

                  & LTM~\cite{PINAYA20221} & \underline{112/171} & 15.12 & {\cellcolor{red!20} 3/18} & {\cellcolor{red!20}6.30} &{\cellcolor{yellow!20}17/26} & {\cellcolor{yellow!20}10.03} & {\cellcolor{yellow!20}11/22} &{\cellcolor{yellow!20}6.66} & {\cellcolor{cgreen!20}\underline{8/10}} & {\cellcolor{cgreen!20}22.95}& {\cellcolor{cgreen!20}\boldmath$18/19$} & {\cellcolor{cgreen!20}33.10} & {\cellcolor{cgreen!20}\boldmath $14/15$} & {\cellcolor{cgreen!20}19.74} & {\cellcolor{red!20}0/1} & {\cellcolor{red!20}0.00} \\
                
                 %%%%% VAE %%%%%
                & VAE~\cite{zimmerer2019unsupervised} &  90/171 & 10.21 & {\cellcolor{red!20}2/18} & {\cellcolor{red!20}4.07} &  {\cellcolor{yellow!20}16/26} & {\cellcolor{yellow!20}13.37} & {\cellcolor{red!20}9/22} & {\cellcolor{red!20}4.90} &  {\cellcolor{cgreen!20}\underline{8/10}} &  {\cellcolor{cgreen!20}16.13}  & {\cellcolor{red!20}7/19} & {\cellcolor{red!20}11.81} & {\cellcolor{cgreen!20}$12/15$} & {\cellcolor{cgreen!20}$14.66$}& {\cellcolor{red!20}0/1}& {\cellcolor{red!20}$0.00$}\\
                %%%%% MAE %%%%%
                & MAE~\cite{he2022masked} & 84/171 & 10.46 & {\cellcolor{red!20}2/18} & {\cellcolor{red!20}4.63} &  {\cellcolor{yellow!20}14/26} & {\cellcolor{yellow!20}13.66} & {\cellcolor{red!20}7/22} & {\cellcolor{red!20}3.71}  &  {\cellcolor{cgreen!20}7/10} &  {\cellcolor{cgreen!20}24.23} & {\cellcolor{red!20}7/19} & {\cellcolor{red!20}17.37} & {\cellcolor{cgreen!20}$12/15$} & {\cellcolor{cgreen!20}$18.48$}  & {\cellcolor{red!20}0/1}& {\cellcolor{red!20}$0.00$}\\
                %%%%% SI-VAE %%%%%
                 & SI-VAE~\cite{daniel2021soft} & 82/171 & 10.01 & {\cellcolor{red!20}0/18} & {\cellcolor{red!20}0.00} & {\cellcolor{red!20}12/26} & {\cellcolor{red!20}7.08} &  {\cellcolor{red!20}6/22} & {\cellcolor{red!20}3.97} &   {\cellcolor{cgreen!20}\underline{8/10}} &  {\cellcolor{cgreen!20}24.55} & {\cellcolor{red!20}9/19} & {\cellcolor{red!20}15.70} & {\cellcolor{cgreen!20}$11/15$} & {\cellcolor{cgreen!20}$14.47$} & {\cellcolor{red!20}0/1}& {\cellcolor{red!20}$0.00$}\\
        	    %%%%% DDPM %%%%%
        	    & DDPM~\cite{Wyatt_2022_CVPR} & 100/171 & 11.03 &  {\cellcolor{red!20}5/18} & {\cellcolor{red!20}7.10} & {\cellcolor{yellow!20}16/26} & {\cellcolor{yellow!20}12.50} & {\cellcolor{red!20}9/22} & {\cellcolor{red!20}4.13} &  {\cellcolor{cgreen!20}7/10} &  {\cellcolor{cgreen!20}13.39} & {\cellcolor{red!20}9/19} & {\cellcolor{red!20}14.89} &  {\cellcolor{cgreen!20}{\boldmath $14/15$}} &  {\cellcolor{cgreen!20}19.39} & {\cellcolor{cgreen!20}\boldmath$1/1$} & {\cellcolor{cgreen!20}5.82}  \\
                \cmidrule{2-18}
                 %%%%% RA %%%%%
                \rowcolor{gray!10} & RA (ours)  & {\boldmath$142/171$} & {\boldmath$39.73$}  & {\cellcolor{cgreen!20}12/18} & {\cellcolor{cgreen!20}\underline{$45.56$}}  & {\cellcolor{cgreen!20} \boldmath $21/26$} & {\cellcolor{cgreen!20} \boldmath $30.78$} & {\cellcolor{cgreen!20}\underline{$17/22$}} & {\cellcolor{cgreen!20}\underline{$29.50$}}  & {\cellcolor{cgreen!20}\boldmath $10/10$} & {\cellcolor{cgreen!20}\boldmath$54.32$}  & {\cellcolor{cgreen!20}\boldmath $18/19$} & {\cellcolor{cgreen!20}\underline{$77.54$}}  & {\cellcolor{cgreen!20}\underline{$13/15$}} & {\cellcolor{cgreen!20}\boldmath$34.78$}  & {\cellcolor{cgreen!20}\boldmath$1/1$} & {\cellcolor{cgreen!20}\underline{$15.38$}}  \\
                
         	    \bottomrule
            \end{tabular}
        \end{adjustbox}
\end{table*}
\begin{figure*}[t!]
    \includegraphics[width=\textwidth]{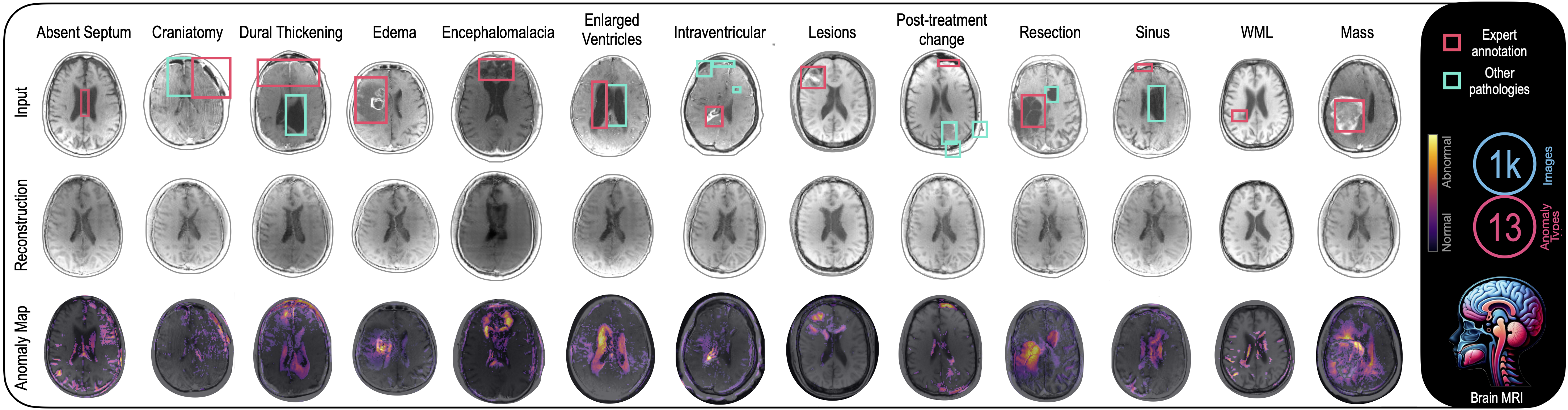}
    % \caption{Anomaly detection on brain MRI. Comprehensive analysis of brain MRI scans using Reversed Auto-Encoders (RA). The top row illustrates the original input images with expert annotations (red) and other pathologies (cyan). The middle row shows pseudo-healthy reconstructions of RA, and the bottom row presents the anomaly maps highlighting detected pathologies in brighter colors. The legend on the right summarizes the composition of the dataset and the diversity of pathologies assessed.\label{fig::adult_brain}}
    \caption{Anomaly Detection in Brain MRI using Reversed Auto-Encoders (RA). The top row displays original brain MRI scans with expert-annotated pathologies (in red) and additional pathologies (in cyan). The middle row depicts the pseudo-healthy reconstructions generated by RA, while the bottom row presents anomaly maps, with detected pathologies highlighted in brighter colors. The legend on the right details the dataset composition and the range of pathologies evaluated. \label{fig::adult_brain}}
\end{figure*}
where \(E_\phi^l\) denotes the \(l\)-th embedding of the \(L\) encoder layers, \(x_{rec} = D_\theta(E_\phi(x))\), \(\mathcal{L}_{Sim}\) is the cosine similarity, and MSE is the mean squared error. The objective function of the encoder, which incorporates the concept of reversed similarity, is defined as:

\begin{equation}
    \mathcal{L}_{E_{\phi}}(x,z) = \text{ELBO}(x) - \frac{1}{\alpha}(\exp(\alpha \text{ELBO}(D_\theta(z))) + \lambda \mathcal{L}_{Reversed}(x),
\end{equation}

\subsection{Anomaly Score Computation}
Beyond reconstruction, accurately detecting anomalies requires a robust anomaly score computation method. Traditional residual-based approaches often face limitations due to their reliance on intensity differences. To address this, we apply adaptive histogram equalization (eq) before computing the residuals. Additionally, we integrate perceptual differences to enhance the robustness of anomaly detection:
\begin{equation}
    s(x) = |\text{eq}(x_{ph}) - \text{eq}(\overline{x})| \times (\mathcal{S}_{\text{lpips}}(x_{ph}, \overline{x}) \times \mathcal{S}_{\text{lpips}}(\text{eq}(x_{ph}), \text{eq}(\overline{x}))),
\end{equation}
where \(\mathcal{S}_{\text{lpips}}\) represents the learned perceptual image patch similarity metric~\citep{zhang2018unreasonable}.

\section{Anomaly Localization on Brain MRI}

Neurological diseases present diverse and complex imaging patterns, ranging from tumors to degenerative diseases. Early and accurate detection of these anomalies is crucial for effective treatment. However, the interpretation of neurological imaging often requires highly specialized expertise, which may not always be readily available. Moreover, the sheer volume and complexity of brain imaging data present significant challenges for manual analysis. This is compounded by the fact that even experienced radiologists can have error rates significant enough to impact patient care. UAD offers a solution by autonomously identifying irregularities in brain imaging, potentially reducing diagnostic errors and improving patient outcomes. This experiment aims to evaluate the effectiveness of our proposed Reversed Auto-Encoders (RA) and various UAD methods in accurately identifying and localizing anomalies across a broad spectrum of brain conditions, thereby underscoring their potential in enhancing neurodiagnostic practices.

\subsection{Datasets}

\textit{Normal Data (Training):} Our training set includes T1-weighted (T1w) MRI scans from FastMRI+~\citep{zhao2021fastmri+}, comprising 131 training, 15 validation, and 30 testing samples, and IXI~(\url{https://brain-development.org/ixi-dataset/}), contributing an additional 581 training samples. These datasets were chosen for their diversity, covering a wide range of normal anatomical variations across different scanners and age groups, to establish a robust normative distribution.

\textit{Pathological Data (Testing):} We utilized the FastMRI+ dataset for its comprehensive annotation of pathologies, including 171 mid-axial T1w slices across 13 distinct pathology classes. This rich dataset facilitates a nuanced assessment of performance, accommodating the diversity of pathological manifestations and the presence of multiple pathologies within single scans, mirroring common clinical challenges.

\subsection{Metrics} The detection performance is evaluated by the number of accurately detected pathologies (\#det) and their precision, measured by the F1 Score. This score represents a balance between precision and recall, with the detailed methodology described in~\cite{bercea2022ra}.
\subsection{Results}
The quantitative evaluation summarized in \autoref{tab::fast}, reveals differing performances corresponding to the complexity of the pathologies.

Denoising Autoencoders (DAEs)~\citep{kascenas2022denoising} have shown commendable results in certain areas like edemas. However, their self-supervised \textit{(Self-S)} nature presents a double-edged sword. As these models are trained to remove or reduce noise from the images, the self-supervised learning process inherently biases the model towards the types of anomalies it has been exposed to during training. This bias can lead to misses of certain types of anomalies that do not fit the learned noise pattern, such as enlarged ventricles or craniotomies. This makes the model less reliable for universal anomaly detection.

Multi-level Knowledge Distillation (MKD)~\citep{salehi2021multiresolution} showed a promising ability to discern anomalies, especially enlarged ventricles, but faced challenges with more complex lesions and craniotomy detection. The visual assessments indicated room for improvement in the precision of their anomaly maps.

Latent Transformer Models (LTMs)~\citep{PINAYA20221} excelled in detecting certain anomalies such as resections and enlarged ventricles but showed limitations with others like edemas. Their performance highlights the potential of likelihood models in medical imaging, especially if combined with more powerful decoders for clearer reconstructions.

Reconstruction-based methods tend to lag in performance compared to other categories. Within this group, Denoising Diffusion Probabilistic Models (DDPM)~\citep{ho2020denoising} stand out as the most proficient, achieving the highest disease identification count with 100 out of 171 total diseases detected.

Reversed Autoencoders (\textit{RA}) emerged as a robust method, consistently delivering detailed and anomaly-free reconstructions across a variety of pathologies, as seen in~\autoref{fig::adult_brain}. \textit{RA} demonstrates a superior ability to balance detection accuracy and precision across a spectrum of pathologies, as evidenced by its leading F1 score of 39.73 and the total of 142 out of 171 detected diseases. 

\section{Anomaly Localization on Pediatric Wrist X-rays}

Wrist injuries, particularly distal radius, and ulna fractures, are prevalent in pediatric patients, often peaking during adolescence. Pediatric surgeons and emergency physicians commonly interpret wrist X-rays, sometimes without the availability of experienced pediatric radiologists. Shortages of radiologists, even in developed nations, can impact patient care, potentially leading to diagnostic errors with error rates as high as 26\%~\citep{nagy2022pediatric}. Detecting anomalies promptly and accurately can expedite treatment decisions and reduce diagnostic errors, ultimately improving outcomes for young patients with wrist injuries. This experiment aims to assess the capability of UAD methods in correctly identifying and localizing various abnormalities in pediatric wrist X-rays.
\begin{table*}[t!]
    \centering 
    \captionof{table}{Anomaly detection and localization results in pediatric wrist X-rays. The table presents Recall and F1 scores for different types of anomalies, including bone anomalies, foreign bodies, fractures, metal objects, periosteal reactions, pronator signs, and soft tissue abnormalities. Our proposed\textit{ Reversed Autoencoders (RA)} demonstrate competitive performance in multiple categories, showcasing their effectiveness in universally detecting and localizing anomalies.\label{tab::wrist}}
    \centering
    \setlength{\tabcolsep}{2pt}
        \begin{adjustbox}{width=\linewidth,center} 
            \begin{tabular}{l l | c c | c c | c c | c c | c c | cc | cc | cc }
                \toprule	    
        
                & \multirow{2}{*}{Method} & \multicolumn{2}{c}{Average} & \multicolumn{2}{c|}{Bone anomaly}  &  \multicolumn{2}{c|}{Foreign body} & \multicolumn{2}{c|}{Fracture} & \multicolumn{2}{c}{Metal} & \multicolumn{2}{c}{Periosteal reaction} & \multicolumn{2}{c}{Pronator sign} & \multicolumn{2}{c}{Soft tissue}\\
                & & Recall $\uparrow$ & F1 $\uparrow$ & Recall $\uparrow$ & F1 $\uparrow$ & Recall $\uparrow$ & F1 $\uparrow$ & Recall $\uparrow$ & F1 $\uparrow$ & Recall $\uparrow$ & F1 $\uparrow$ & Recall $\uparrow$ & F1 $\uparrow$ & Recall $\uparrow$ & F1 $\uparrow$ & Recall $\uparrow$ & F1 $\uparrow$\\\midrule
                \textit{Self-S} & DAE~\cite{kascenas2022denoising} & \underline{60.42} & \underline{15.71} & 43.33 & {11.05} & \underline{75.00} & \underline{25.38} & \underline{59.72} & {\boldmath $15.19$} & \underline{$97.10$} & 41.86 & {59.18} & \underline{13.65} & 33.33 & 7.4 & \underline{$28.94$} & \underline{9.73} \\ \midrule
                 % \multirow{7}{*}{\rotatebox[]{90}{\textit{Unsupervised}}} & MKD~\cite{salehi2021multiresolution} &\\
                 \multirow{6}{*}{\rotatebox[]{90}{\textit{Unsupervised}}}  & LTM~\cite{PINAYA20221} & 30.16 & 6.99 & 40.56 & 9.43 & \underline{75.00} & 14.58 & 25.17 & 4.85 & 90.33 & {\boldmath $55.56$} & 42.36 & 7.66 & \underline{66.67} & 11.61 & 15.79 & 5.62 \\
        	    & VAE~\cite{zimmerer2019unsupervised} & 42.14 & 9.15 & {43.88} & 10.33  & \underline{75.00} & 13.33 & 38.58 & 7.23 & 90.09 & \underline{ $54.82$} & 51.23 & 9.54 & \underline{ $66.67$} & 11.11 &7.89 & 3.49\\
                & MAE~\cite{he2022masked} & 23.80 & 5.87 & 22.78 & 5.66  & \underline{75.00} & {\boldmath $33.33$} & 19.44 &4.06 &77.78 & 45.18 & 34.97 &6.89 & 100 & \underline{17.77} & 10.52 & 4.93 \\
         	    & SI-VAE~\cite{daniel2021soft}  & 49.56 & 10.03 & 42.22 & 8.99  & \underline{75.00} & 25.0 & 46.86 & 8.52 & 92.51 & 45.47 & 56.09 & 10.60 & {\boldmath $100$} & {\boldmath$19.44$} & 15.79 & 4.90 \\ 
                & DDPM~\cite{Wyatt_2022_CVPR} & 55.15 & 11.25 & \underline{$58.33$} & \underline{12.49}  & {\boldmath $100$} & 19.64  & 50.99 & 9.41 & {\boldmath $97.34$} & 50.70 & {\boldmath $68.07$} & 12.41 & {\boldmath $100$} & 16.66 & {$25.00$} & 7.59  \\ \cmidrule{2-18}
                \rowcolor{gray!10} & RA (ours) & {\boldmath $65.26$} & {\boldmath $16.11$} & {\boldmath$65.56$} &  {\boldmath $22.65$}  & \underline{75.00} & {$18.75$} & {\boldmath $65.20$} & \underline{$14.40$} & {96.86} &  {$46.93$}  & \underline{ $60.36$} &  {\boldmath $17.99$} &  \underline{ $66.67$} & {$15.00$} & {\boldmath $31.58$} & {\boldmath $11.26$}\\ % & 90.27 & 58.96 \\
         	    \bottomrule
            \end{tabular}
        \end{adjustbox}
\end{table*}
\begin{figure*}[t!]
    \includegraphics[width=\linewidth]{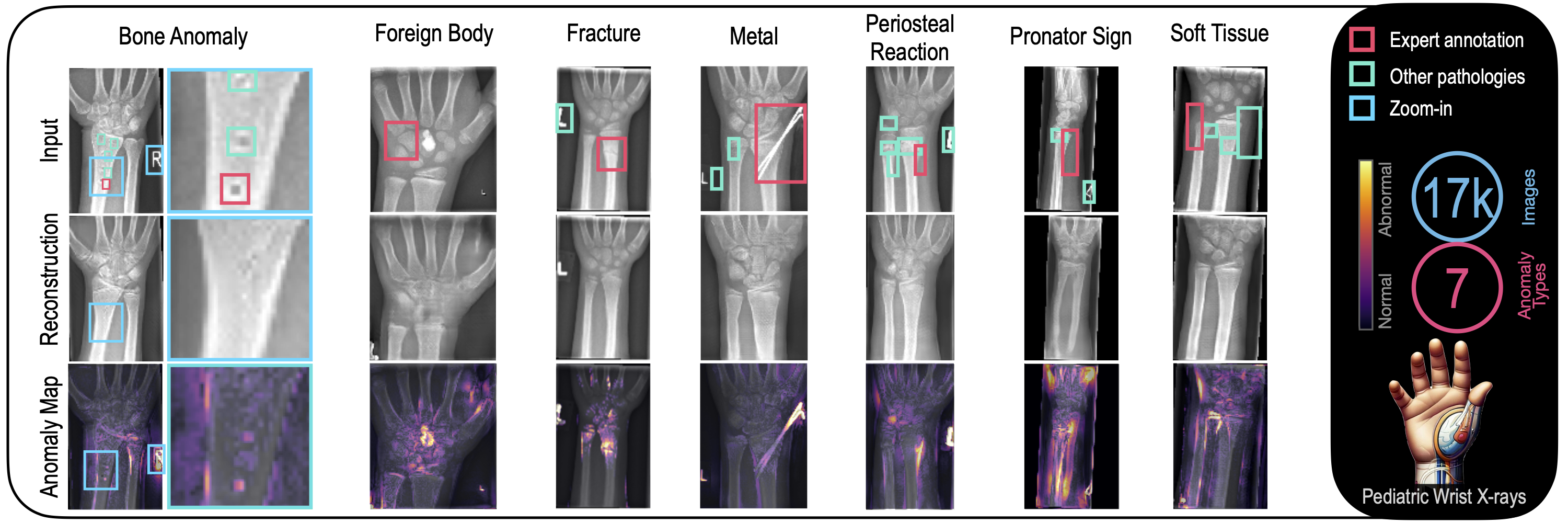}
    \caption{Anomaly detection on pediatric wrist X-rays showcasing a comparison between the original input images, the reconstructed images using our method \textit{(RA)}, and the corresponding anomaly maps. Each column represents a different category of anomaly identified in the study, with expert annotations evaluated, other present pathologies, and areas of interest for zoom-in highlighted. The anomaly maps are color-coded to facilitate the localization and visualization of potential pathologies. The dataset encompasses 17k images and 7 anomaly types, demonstrating the diversity and complexity of the clinical conditions analyzed.\label{fig::pediatric_wrist}}
\end{figure*}
\subsection{Dataset}
The dataset utilized in this experiment, known as GRAZPEDWRI-DX~\citep{nagy2022pediatric}, comprises 10,643 pediatric wrist radiography studies from 6,091 unique patients, with a mean age of 10.9 years. It encompasses various abnormalities, including fractures, metal implants, periosteal reactions, bone lesions, soft tissue swelling, osteopenia, plaster casts, and the pronator quadratus sign. Annotations by board-certified pediatric radiologists include bounding boxes.

\subsection{Metrics} 
Unlike supervised object detection methods, unsupervised anomaly detection methods do not provide bounding boxes or explicit object localization information. As a result, traditional metrics such as Intersection over Union (IoU) or overlap calculations cannot be computed in the context of unsupervised anomaly detection. Instead, we rely on alternative metrics such as the number of accurately detected pathologies (\#det) and the F1 Score, which assess the quality of detections without the need for bounding boxes~\citep{bercea2022ra}. 

\subsection{Results}

The comparative analysis of anomaly detection methods in pediatric wrist X-rays, presented in \autoref{tab::wrist} and illustrated in \autoref{fig::pediatric_wrist}, provides a detailed overview of the capabilities and limitations of each approach.

Denoising Autoencoders (DAE), while adept at highlighting areas of increased density potentially indicative of inflammation, frequently missed the primary fractures. The occurrence of such hyperintensities in X-rays is typically a response to bone injury, where physiological changes like inflammation lead to local increased blood flow and fluid accumulation. These changes result in radiographic hyperdensity, which DAEs are prone to detect. However, their emphasis on these secondary signs without direct fracture visualization underscores a critical limitation—failing to identify the essential diagnostic feature of the fracture itself.

Conversely, while methods like DDPM showed effectiveness in specific categories of anomalies, their performance was not uniform across all areas, often displaying significantly lower precision as indicated by the F1 score. This uneven performance highlights the fundamental difficulties in developing an unsupervised anomaly detection system that is consistently proficient in all aspects.

\textit{RA} manifested a more uniformly competitive performance, particularly in identifying fractures and soft tissue abnormalities, as evidenced by their high recall and F1 scores. Nonetheless, RA faced challenges in the detection of very subtle anomalies amid the variability inherent in the normal bone structures of pediatric patients. Such difficulties are exemplified in the 'Bone Anomaly' section of \autoref{fig::pediatric_wrist}, demonstrating a struggle of unsupervised methods with small lesions. This struggle is exacerbated by traditional evaluation metrics like the Dice coefficient or bounding box overlap, which may not effectively capture the subtleties of anomaly maps when pathologies present minimally against the complex anatomy of developing bones.

The findings underscore the imperative for more advanced anomaly map computations and the adoption of evaluation metrics attuned to the intricacies of unsupervised anomaly detection, to better support the clinical decision-making process.
\section{Chest X-ray Anomaly Detection}

Chest radiographs are an essential diagnostic tool in identifying respiratory conditions such as pneumonia. However, distinguishing normal findings from those indicative of pathological conditions can be challenging due to overlapping imaging features. In the context of the COVID-19 pandemic, the need for efficient and accurate diagnostic methods has become even more pressing. Traditional diagnostic approaches rely heavily on the expertise of radiologists, who face an increased workload and the risk of diagnostic errors, especially during peak times of respiratory illnesses.

UAD presents a promising solution to these challenges. It offers the capability to autonomously detect subtle and complex patterns indicative of respiratory diseases, potentially enhancing diagnostic accuracy and speed. Our experiment is designed to assess the ability of \textit{RA} and other UAD methods to accurately differentiate normal chest radiographs from those showing anomalies indicative of pneumonia and COVID-19. The goal is to assess the precision of these AI-driven methods in identifying specific anomalies associated with each condition, thereby contributing to more efficient and accurate clinical decision-making in respiratory care.
\begin{table*}[t!]
    \centering 
    \captionof{table}{CXR Pathology Detection using \textit{Reversed Auto-Encoders (RA). RA} excels in generating accurate pseudo-healthy reconstructions of chest X-rays (CXR), crucial for the precise localization and identification of pathologies. The best results are highlighted in \textbf{bold}, and the second-best results are \underline{underlined}. For a detailed visual representation of the pathology detection and localization capabilities of \textit{RA}, see~\autoref{fig::cxr}.
    \label{tab::benchmark_sota_anomaly}}

    % \captionof{table}{CXR Pathology Detection. \textit{RA} generates accurate pseudo-healthy reconstructions which enables the precise localization of pathologies. See visual results in~\autoref{fig::cxr}. 

    \centering
    \setlength{\tabcolsep}{2pt}
        \begin{adjustbox}{width=\linewidth,left} 
            \begin{tabular}{l l| c c | c c c c | c c c c}
                \toprule	    
                & \multirow{3}{*}{Method} & \multicolumn{2}{c|}{Healthy}  & \multicolumn{4}{c}{Pneumonia}& \multicolumn{4}{c}{Covid-19} \\
                & & SSIM $\uparrow$ & LPIPS $\downarrow$ & {AUROC $\uparrow$} & {AUPRC $\uparrow$} & {$FP\%TP95$} $\downarrow$ & {$FP\%TP99 \downarrow$}  & {AUROC $\uparrow$} & {AUPRC $\uparrow$} & {$FP\%TP95$} $\downarrow$ & {$FP\%TP99$} $\downarrow$ \\\midrule
                \textit{Self-S} & DAE~\cite{kascenas2022denoising}  & {\boldmath $96.3^*$} & {\boldmath $1.2^*$}  &  \underline{$82.6$} & \underline{95.81} & \underline {$57.41$} & \underline{86.46} & \underline{$78.9$} & \underline{87.53} & \underline{73.50} & \underline{92.91} \\ \hline
        	    \multirow{7}{*}{\rotatebox[]{90}{\textit{Unsupervised}}} & MKD~\cite{salehi2021multiresolution} & N/A & N/A & 27.22 & 76.28 & 98.72 & 99.80 & 36.05 & 57.87 & 98.83 & 99.84\\ 
             & LTM~\cite{PINAYA20221} & \underline{75.30} & 22.36 & 58.05 & 88.63 &91.76 & 97.35 & 62.03 & 76.59 & 92.75 & 98.60\\
              & VAE~\cite{zimmerer2019unsupervised} & 75.22 & 31.89 & 40.49 & 83.21 & 97.15 & 99.41 & 50.29 & 68.22 & 98.67 & 100  \\
              & MAE~\cite{he2022masked} & 71.35 & 34.09 & 70.40 & 93.28 & 90.28  & 96.37 &  64.89 & 78.38 & 93.53 & 99.14\\
         	   & SI-VAE~\cite{daniel2021soft}  & {69.36} & 11.78 & 52.97 & 87.68 & 95.19 & 97.94 & 58.67 & 75.19 & 95.64 & 99.14 \\
               & DDPM~\cite{ho2020denoising} &  68.03 & 9.95 & $54.29$  & 88.30 & 94.31 & 98.23 & $48.42$ & 68.26 & 98.05 & 99.77 \\  
               \rowcolor{gray!10} &  RA (ours) & {67.37} & \underline{9.93}  & {\boldmath $84.64$}  & {\boldmath $96.52$} & {\boldmath $53.39$} & {\boldmath $82.04$} &  {\boldmath $84.69$} & {\boldmath $91.70$} & {\boldmath $68.82$} & {\boldmath $89.32$} \\
         	    \bottomrule
            \end{tabular}
        \end{adjustbox}
\end{table*}

\subsection{Datasets}
The RSNA dataset~\citep{rsna}, consisting of 10,000 normal and 6,000 lung opacity CXR images, was used to represent a range of pathological conditions. The Padchest dataset~\citep{bustos2020padchest} was employed for COVID-19 detection, comprising 1,300 normal control images and 2,500 COVID-19 cases. All images were standardized to a resolution of $128\times128$ pixels.

\subsection{Metrics}
This study assesses anomaly detection in chest X-ray images using various metrics. For healthy cases, SSIM (Structural Similarity Index Measure) and LPIPS~\citep{zhang2018unreasonable} (Learned Perceptual Image Patch Similarity) are used. For anomalies, the methods are evaluated based on AUROC (Area Under the Receiver Operating Characteristic curve), AUPRC (Area Under the Precision-Recall Curve), and False Positives at True Positive rates of 95\% (FP@TP95) and 99\% (FP@TP99). 

\subsection{Results}
\begin{figure}[t!]
    \includegraphics[width=\columnwidth]{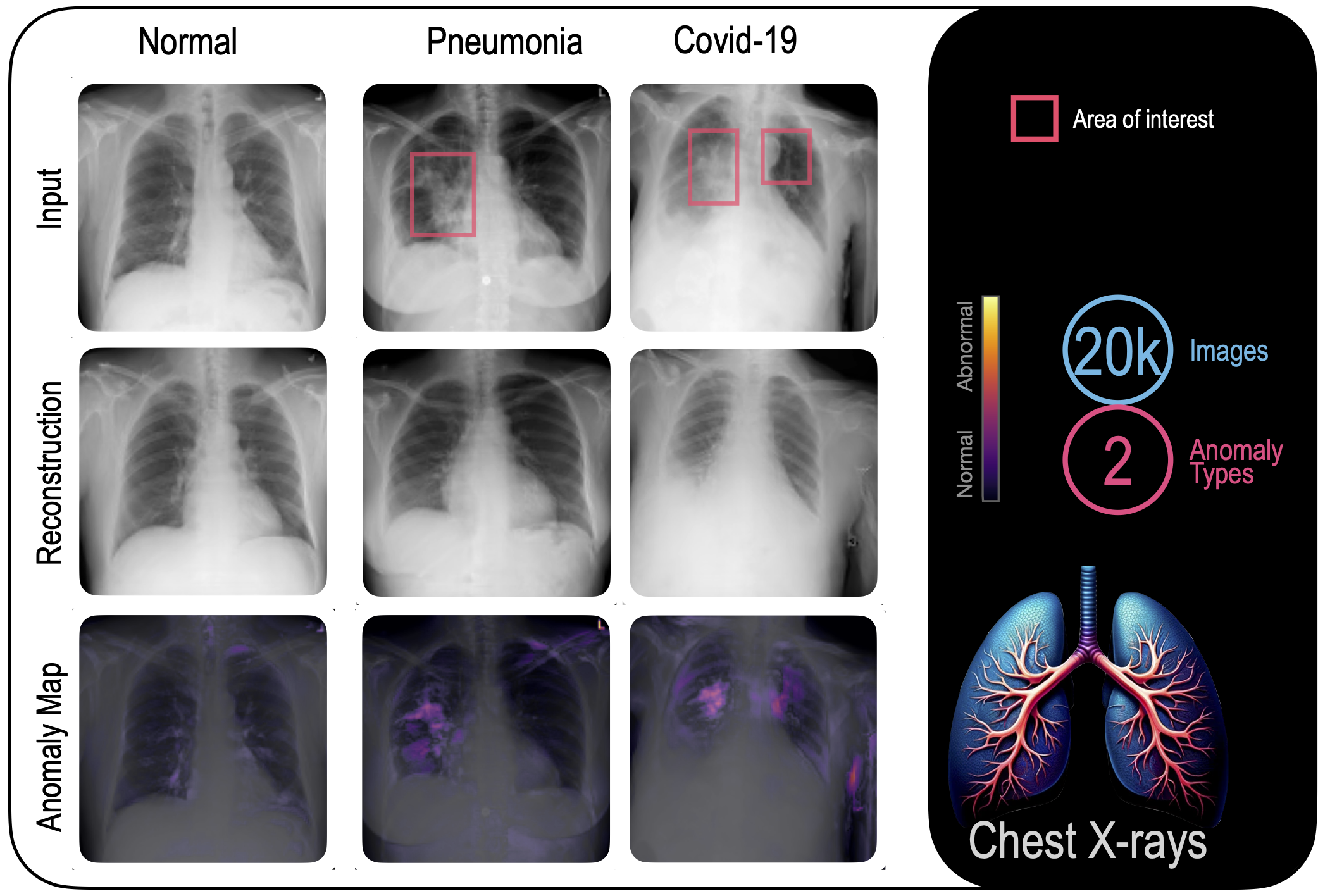}
    \caption{Anomaly detection on chest X-rays. The figure illustrates a comparison across three panels: normal, pneumonia, and COVID-19 CXRs. For each condition, the top row presents the original input images, the middle row shows the pseudo-healthy reconstructions and the bottom row displays the corresponding anomaly maps. Anomalies are indicated by red boxes in the input images and are highlighted in the anomaly maps to indicate the severity and location of the pathology. The dataset comprises 20k images spanning two anomaly classes. \label{fig::cxr}}
\end{figure}

~\autoref{fig::cxr} demonstrates the capability of \textit{RA} to generate pseudo-healthy reconstructions with corresponding anomaly maps that accentuate regions of pathology. Compared to other UAD methods, \textit{RA} achieved the highest AUROC scores for identifying pneumonia and COVID-19, as detailed in~\autoref{tab::benchmark_sota_anomaly}. These results highlight the potential of \textit{RA} in clinical settings for accurately detecting and localizing lung pathologies in CXR images, underscoring its suitability for incorporation into diagnostic workflows.
\section{Discussion}

In this study, we introduced the \textit{Reversed Auto-Encoders (RA)}, an unsupervised anomaly detection framework, and conducted an extensive evaluation across various medical imaging modalities. The ability of \textit{RA} to generate pseudo-healthy reconstructions contributes to addressing a significant challenge in medical imaging analysis: the unbiased detection of pathologies.

The potential clinical value of \textit{RA} lies in its autonomous anomaly detection capability, especially beneficial in environments with scarce radiological expertise. We have tested its versatility and robustness on diverse datasets, including brain MRI, pediatric wrist X-rays, and chest X-rays. \textit{RA} has exhibited proficiency in detecting subtle anomalies amid the complex variability of normal anatomical structures, indicating an improvement over existing methodologies.

However, our research also sheds light on the limitations in detecting extremely subtle anomalies. The analysis of pediatric wrist X-rays, for example, highlights the necessity for refining anomaly map computations and developing more sophisticated evaluation metrics tailored to the intricate demands of clinical diagnostics.

Our findings underline the importance of comprehensive evaluations of anomaly detection methods across varied pathologies and anatomical contexts. Such extensive benchmarking is crucial for the transition of these methods from research to clinical application, revealing current limitations and guiding future research directions.

To summarize, the \textit{RA} framework demonstrates promising potential in medical imaging. Its generalized ability to detect a wide range of anomalies with notable accuracy contributes meaningfully to the field. This work advances the intersection of medical imaging and artificial intelligence, offering clinically relevant insights that could improve diagnostic processes. While it represents a step towards automated, precise, and universally applicable diagnostic tools, continued research and development are essential to fully realize these objectives and enhance the support they offer to medical practitioners and patient care.


%%Harvard
\bibliographystyle{model2-names.bst}\biboptions{authoryear}
\bibliography{main}

\end{document}